\documentclass[aps,prl,shownopacs,superscriptaddress,twocolumn]{revtex4}

\usepackage{siunitx}
\usepackage{bm}
\usepackage{ifpdf}
\usepackage{float}
\usepackage{upgreek}
\usepackage{xcolor}
\usepackage{graphicx}
\usepackage{hyperref}
\usepackage{amsmath}
\usepackage{epstopdf}
\usepackage{bbm}
\begin{document}

\title{Experimental Fock-State Bunching Capability of Non-Ideal Single-Photon States}

\author{Petr Zapletal\footnotemark[3]\footnotetext{\footnotemark[3]Present address: Friedrich-Alexander University Erlangen-N\"urnberg (FAU), Department of Physics, 91058 Erlangen, Germany.}}
\affiliation{Department of Optics, Palack\'y University, 17. listopadu 1192/12, 77146 Olomouc, Czech Republic}
\author{Tom Darras}
\affiliation{Laboratoire Kastler Brossel, Sorbonne Universit\'e, CNRS, ENS-Universit\'e PSL, Coll\`ege de France, 4 Place
Jussieu, 75005 Paris, France}
\author{Hanna Le Jeannic \footnotemark[4]\footnotetext{\footnotemark[4]Present address: Laboratoire Photonique Num\'erique et Nanoscience, Universit\'e de Bordeaux, Institut d'Optique, CNRS, UMR 5298, 33400 Talence, France.}}
\affiliation{Laboratoire Kastler Brossel, Sorbonne Universit\'e, CNRS, ENS-Universit\'e PSL, Coll\`ege de France, 4 Place
Jussieu, 75005 Paris, France}
\author{Adrien~Cavaill\`{e}s}
\affiliation{Laboratoire Kastler Brossel, Sorbonne Universit\'e, CNRS, ENS-Universit\'e PSL, Coll\`ege de France, 4 Place
Jussieu, 75005 Paris, France}
\author{Giovanni Guccione}
\affiliation{Laboratoire Kastler Brossel, Sorbonne Universit\'e, CNRS, ENS-Universit\'e PSL, Coll\`ege de France, 4 Place
Jussieu, 75005 Paris, France}
\author{Julien Laurat}
\email{julien.laurat@sorbonne-universite.fr}
\affiliation{Laboratoire Kastler Brossel, Sorbonne Universit\'e, CNRS, ENS-Universit\'e PSL, Coll\`ege de France, 4 Place
Jussieu, 75005 Paris, France}
\author{Radim Filip}
\email{filip@optics.upol.cz}
\affiliation{Department of Optics, Palack\'y University, 17. listopadu 1192/12, 77146 Olomouc, Czech Republic}

\begin{abstract}
Advanced quantum technologies, as well as fundamental tests of quantum physics, crucially require the interference of multiple single photons in linear-optics circuits. This interference can result in the bunching of photons into higher Fock states, leading to a complex bosonic behaviour. These challenging tasks timely require to develop collective criteria to benchmark many independent initial resources. Here we determine whether $n$ independent imperfect single photons can ultimately bunch into the Fock state $|n \rangle$. We thereby introduce an experimental \textit{Fock-state bunching capability} for single-photon sources, which uses phase-space interference for extreme bunching events as a quantifier. In contrast to autocorrelation functions, this operational approach takes into account not only residual multi-photon components but also vacuum admixture and the dispersion of the individual photon statistics. We apply this approach to high-purity single photons generated from an optical parametric oscillator and show that they can lead to a Fock-state capability of at least $14$. Our work demonstrates a novel collective benchmark for single-photon sources and their use in subsequent stringent applications.
 \end{abstract}
 
 \maketitle

\section{Introduction}
Beyond its fundamental significance, the Hong-Ou-Mandel (HOM) effect \cite{HOM}, where two single photons interfere on a beamsplitter, has been central to the development of quantum technologies. With the advance of complex quantum information protocols and networking architectures \cite{Walmsley,flamini2019}, the availability of multiple indistinguishable photons is becoming a cornerstone. Multi-photon interference is required in quantum processing with optical \cite{peruzzo2010, crespi2013, tillmann2013, broome2013, spring2013, spagnolo2014, bentivegna2015, wang2017,zhong2018,Minzioni2019,Guccione2020} or microwave photons \cite{Pfaff2017,gao2018,Peropadre2016}, ranging from boson sampling studies to quantum state engineering. It also plays a key role in quantum sensing \cite{matthews2016,ulanov2016,slussarenko2017}, noiseless amplification \cite{ralph2009}, quantum key distribution \cite{ghalaii2020a,ghalaii2020b} or error correction \cite{michael2016,bergmann2016,hu2019}.

\begin{figure*}[htpb!]
\vspace{-0.6cm}
\centering
\includegraphics[width=1.7\columnwidth]{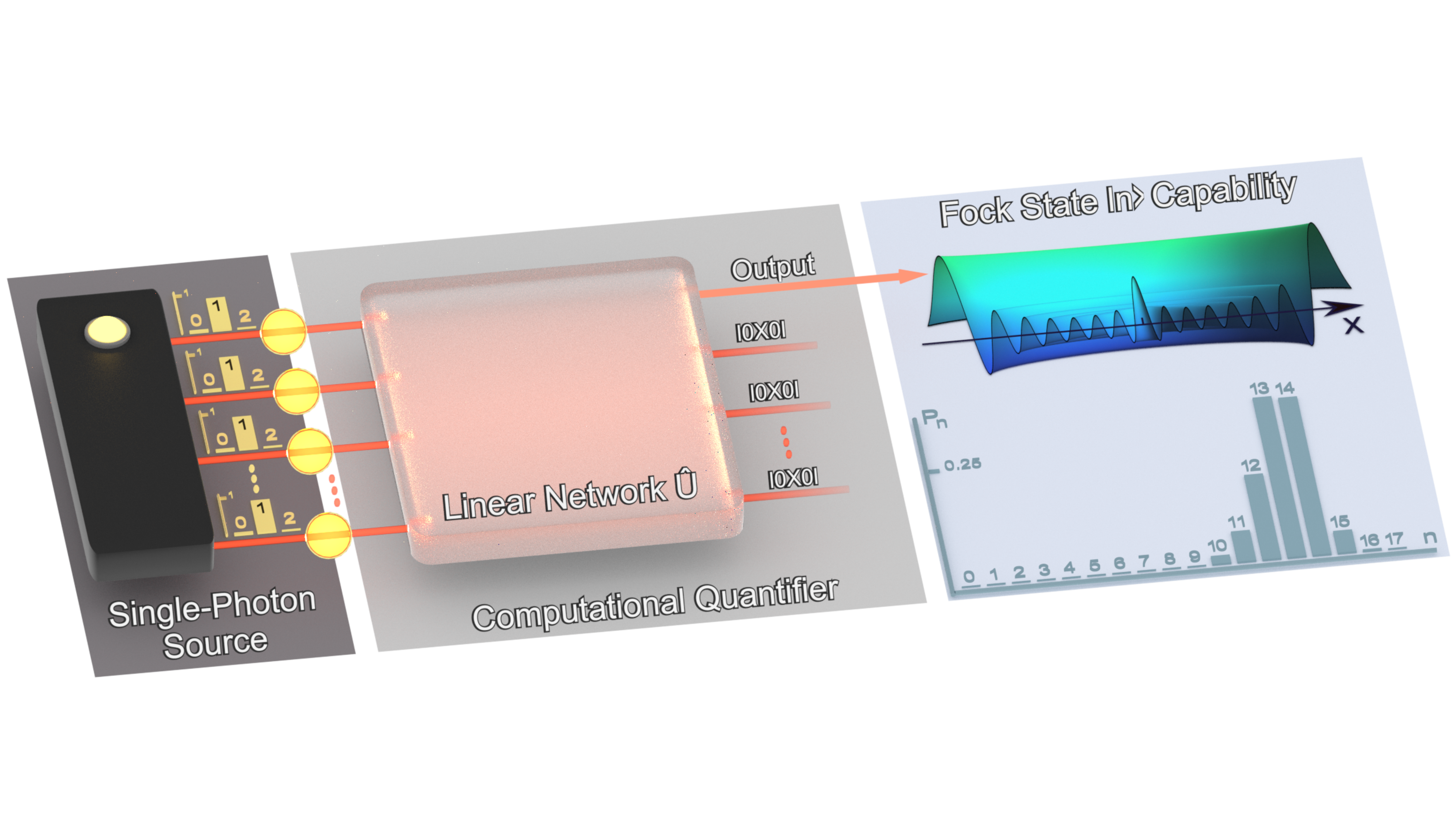}
\vspace{-1.3cm}
\caption{Fock-state bunching capability of non-ideal single-photon states. A single-photon source provides photons with different vacuum admixture and residual multi-photon components, as depicted by the photon-number distributions (left). These states are used as inputs of a balanced linear optics network $\hat{U}$. In an extreme case, all photons can bunch into just one output mode whereas all other modes are in the vacuum state. This stage is done computationally and provides the expected photon-number distribution $P_n$ for the output mode (right). The negativities of the associated Wigner function are used to determine the \textit{Fock-state capability}. In contrast to other measures, this collective bechmark depends not only on the vacuum admixture and multiple-photon statistics of the imperfect input photons but also on the small discrepancies between them.}
\label{scheme}
\end{figure*}

Multi-photon interference leads to a non-trivial redistribution of photons between optical modes. To achieve such interferences, all photons have to be indistinguishable. Several methods have been recently developed to investigate this indistinguishability using different benchmarks, e.g., fidelity \cite{aolita2015} or specific photon correlation measures \cite{crespi2016,viggianiello2018,giordani2018,brod2019,flamini2020}. However, the joint impact of photon statistics from many imperfect single-photon states, i.e., exhibiting unwanted vacuum and residual multi-photon components, on multi-photon interference has remained elusive. The joint statistical influence of these parameters cannot be described by evaluating properties of single-photon states that are averaged over many experimental runs. Hence, we need criteria, experimental data and subsequent analysis to determine whether independently generated single photons can, in principle, produce the targeted multi-photon interference effects. 

Multi-photon interference effects come in a variety of flavors. An extreme event corresponds to the bunching of $n$ single photons into the Fock state $|n\rangle$ \cite{steuernagel1997,escher2005,motes2016}. Such bunching can appear in a linear-optical network with inputs fed by indistinguishable single photons, as shown in Fig.~\ref{scheme}. The elementary example is the appearance of Fock state $|2\rangle$ based on the HOM effect, as demonstrated in experiments with optical photons but also with microwave photons \cite{lang2013,gao2018}, phonons in trapped ions \cite{toyoda2015} or surface plasmons \cite{Zwiller,Atwater}. This extreme bunching event, i.e., the result of a clear operational procedure, enables to introduce a strong benchmark for single-photon states that evaluates their ability to undergo multi-photon interference \cite{zapletal2017}. This \textit{Fock-state bunching capability} relies on negativities of the resulting Wigner function, which provide a very sensitive signature of the non-classicality of the generated higher Fock states \cite{schleich2001,HarocheBook}.

In contrast to the well-known second-order autocorrelation function at zero time delay $g^{(2)}(0)$, which measures the suppression of the multi-photon contribution and affects the interference visibility \cite{Ollivier2020}, the capability is also strongly dependent on the vacuum admixture. Another crucial difference is that it collectively tests multiple photon statistics and determines the joint statistical impact of small discrepancies between them. This provides more stringent and accurate evaluation than other available characteristics.

The previous theoretical study based on Monte-Carlo simulations has only predicted that the bunching of single photons is affected by vacuum and multi-photon contributions \cite{zapletal2017}. However, these contributions and their dispersion in non-ideal photon statistics of many independent copies are too complex to be described, specifically when the number of photons increases. Experimental data are necessary to confirm this prediction. Here, we employ the bunching capability to collectively benchmark experimental single-photon states using heralded single photons generated by parametric down-conversion from an optical parametric oscillator (OPO). By tuning the photon source properties, we address the scaling of the capability with the statistics of non-ideal single photons. We hereby provide a crucial insight into the combined effects of non-ideal photon statistics of independently generated single photons. We demonstrate that experimentally generated single photons can bunch into the Fock state $|14\rangle$ with high fidelity and suppressed higher Fock states contributions. We show that the Fock state capability non-linearly decreases with photon loss, providing a more stringent characterization than $g^{(2)}(0)$, which is independent of photon loss, and also than negative Wigner function that decreases only linearly. Our results indicate that despite the negative impact of multi-photon contributions typically reported using $g^{(2)}(0)$, they prevent the bunching of single-photon states into a respective Fock state less severely than optical loss.

\section{Quantifier principle}
We first describe the quantifier principle. To collectively test the ability of the generated single photons to undergo multi-photon interference, we computationally determine the Wigner function of the higher Fock state, which can, in principle, appear from multiple copies of the single-photon state, as depicted in Fig.~\ref{scheme}. The area in phase space, where the Wigner function of the ideal Fock state $|n\rangle$ is negative, is composed of $n/2$ or $(n-1)/2$ concentric annuli if $n$ is an even number or an odd number, respectively. By definition, a single-photon state has the capability of the Fock state $|n\rangle$ if the Wigner function of the state, which can be generated from $n$ independent copies of the single-photon state, has the same number of negative annuli as the ideal Fock state $|n\rangle$ \cite{zapletal2017}. The negative annuli in the Wigner function witness the non-classical nature of the multi-photon interference in phase space. The Fock-state capability, which is determined computationally, collectively tests the copies of a single-photon state, even though any multi-copy procedure is not implemented in the laboratory. 

In theory, copies of the ideal single-photon state $|1\rangle$ have the capability of an arbitrary Fock state $|n\rangle$. For states generated by single-photon sources, the negative annuli in the Wigner function are sensitive to the presence of vacuum and multi-photon contributions. Also, the exact distribution of residual multi-photon statistics in many non-ideal single-photon states is not known. As a consequence, the joint effect of small discrepancies between individual single-photon copies on multi-photon interference has to be investigated by applying the quantifier on photon statistics measured in an experiment. In this way, we can determine whether the single-photon sources have a sufficient quality for applications in quantum technology that require multi-photon interference.

\section{Single-photon generation and multiple data sets}
To study this benchmark, we used heralded single-photon states generated using a two-mode squeezer, i.e., a type-II phase-matched optical parametric oscillator operated well below threshold (see Appendix). The signal and idler photons at 1064 nm are separated on a polarizing beam-splitter and the idler photon is detected via a high-efficiency superconducting nanowire single-photon detector. This detection event heralds the generation of a single photon in the signal mode. The generated state is emitted into a well-defined spatio-temporal mode \cite{Morin2013}, with a bandwidth of about 65 MHz. The state is measured via high-efficiency homodyne detection, with a visibility of the interference with the local oscillator above 99\%, and reconstructed via maximum-likelihood algorithms \cite{maxlik}. The experimental setup has been described elsewhere \cite{Morin2013b,LeJeannic2016}. 

Importantly, the OPO used in this work exhibits a close-to-unity escape efficiency, i.e., the transmission of the output coupler is much larger than the intracavity losses \cite{Morin2014}. As a result, a large heralding efficiency can be obtained, i.e., a very low admixture of vacuum. A single-photon component up to 91\% is achieved. Also, by changing the pump power the multi-photon component can be increased at will. These features enable us to explore different combinations of state imperfections. Seven sets of data were recorded, each of them being obtained by a repetitive measurement of the single-photon states generated under the same conditions. Parameters of the sets are given in Table \ref{table}. They include the single-photon component $P_1$ and the probability $P_{2+}$ of finding two or more photons. These measured quantities give also access to the conditional second-order autocorrelation function at zero-time delay $g^{(2)}(0)$ \cite{LeJeannic2016}.

\setlength{\tabcolsep}{1mm}{
\begin{table}[t!] 
 \caption{Photon-number statistics of heralded single photons. Each set is obtained by successive measurements under the same conditions (in particular pump power). The table displays the single-photon component $P_1$, the multi-photon probability $P_{2+}$, the second-order correlation function $g^{(2)}(0)$, and the negativity at the origin of the Wigner function.} 
\vspace{0.2cm}
\centering
 \begin{tabular}{| c |c | c | c  |c|}
            \hline
 		& $P_1$ & $P_{2+}$ &$g^{(2)}(0)$&$2\pi\times W(0,0)$\\
	   \hline
 $1$ & $0.53\pm0.01$ &$0.010\pm0.006$ & $0.07\pm0.05$& $-(0.05 \pm 0.02)$   \\   	   
 $2$ & $0.62\pm0.02$ &$0.013\pm0.008$ & $0.07\pm0.04$& $- (0.23 \pm 0.03)$\\ 
 $3$ & $0.74\pm0.01$ & $ 0.016\pm0.008$  & $0.06\pm0.03$& $- (0.47 \pm 0.02) $  \\
 $4$ & $0.72\pm0.01$ &  $0.05\pm0.01$  & $0.2\pm0.04$& $- (0.45 \pm 0.02) $  \\
 $5$ & $0.83\pm 0.01$ & $0.07\pm 0.01$  &$0.2\pm 0.03$& $- (0.67 \pm 0.02)$ \\
  $6$& $0.86\pm 0.01$ & $0.02\pm 0.01$  &$0.05\pm 0.03$& $- (0.73 \pm 0.02) $   \\	   
  $7$ &  $0.91\pm 0.01$ & $0.02\pm 0.01$ & $0.05\pm 0.02$& $- (0.82 \pm 0.02) $ \\
             \hline
 \end{tabular} 
 \label{table}
\end{table}}

\begin{figure}[b!]
\centering
\includegraphics[width=0.9\columnwidth]{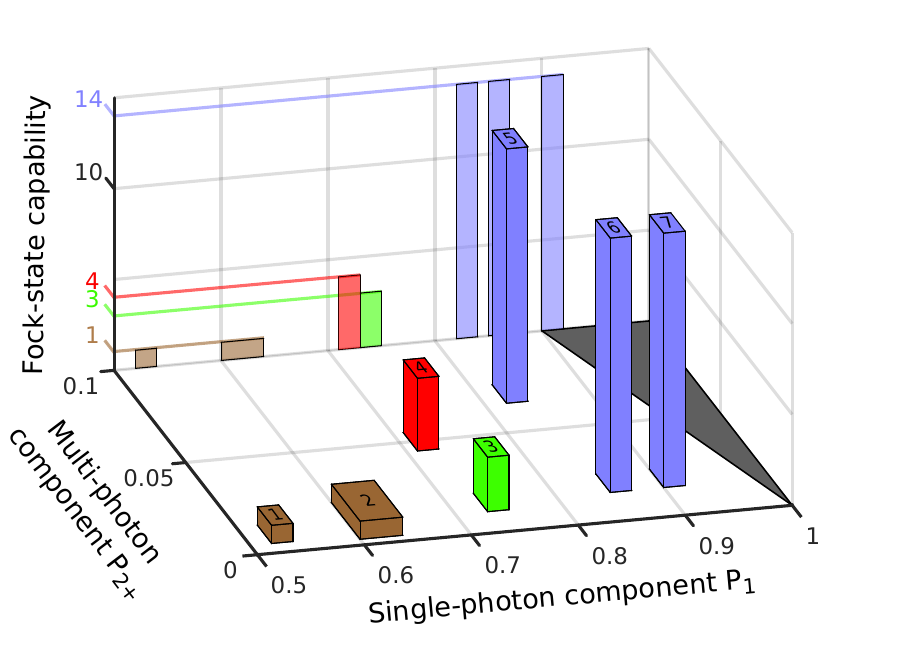}
\vspace{-0.2cm}
 \caption{Fock-state bunching capability of the experimentally-generated single-photon states. The capability is given as a function of the single-photon component $P_1$ and of the multi-photon probability $P_{2+}$ for the different sets given in Table I. These parameters are averaged over photon-number statistics from a given data set obtained by successive measurements under the same experimental conditions. Colors denote the Fock-state capability. The gray-shaded area excludes the unphysical probabilities $P_1+P_{2+}>1$. The standard deviation of the probabilities are given by the thicknesses of the color bars.}
 \label{cap}
\end{figure}

\section{Experimental Fock-state capability}
To test a particular data set for the Fock-state capability $n$, the data are randomly partitioned in $n$ subsets from which $n$ photon-number statistics are obtained and used as the quantifier inputs. The output-state Wigner function of the computational quantifier is averaged over 30 such random choices. From the averaged output-state Wigner function, it is determined whether the data set has the Fock-state capability $n$ (see Appendix). The capability for all data sets is depicted in Fig.~\ref{cap} as a function of $P_1$ and $P_{2+}$. The quantifier is presently computationally limited by the Fock-state capability $14$ (see Appendix), which is already a very large number in this operational context. All data sets for which this capability 14 is obtained may also have the capability of a higher Fock state. In the following, we describe the different measured points and typical trends.

%bad states
First, single-photon states with a low purity due to a vacuum component close to 50\% (brown bars in Fig.~\ref{cap}, sets 1 and 2 in Table \ref{table}) have only the trivial capability of the Fock state $|1\rangle$, despite their very low $g^{(2)}(0)$. This shows that the broadly used autocorrelation function does not fully characterize the ability to bunch into higher Fock states exhibiting non-classical signatures. In particular this example demonstrates that the capability is more sensitive to vacuum mixture, as a state obtained from two copies of these single photons would have a positive Wigner function. Due to their trivial capability, such states are not a useful resource for the preparation of large Fock states that could be used e.g. for quantum metrology \cite{matthews2016,ulanov2016,slussarenko2017} or error correction \cite{michael2016,bergmann2016,hu2019}.

\begin{figure}[t!]
\centering
\includegraphics[width=0.85\columnwidth]{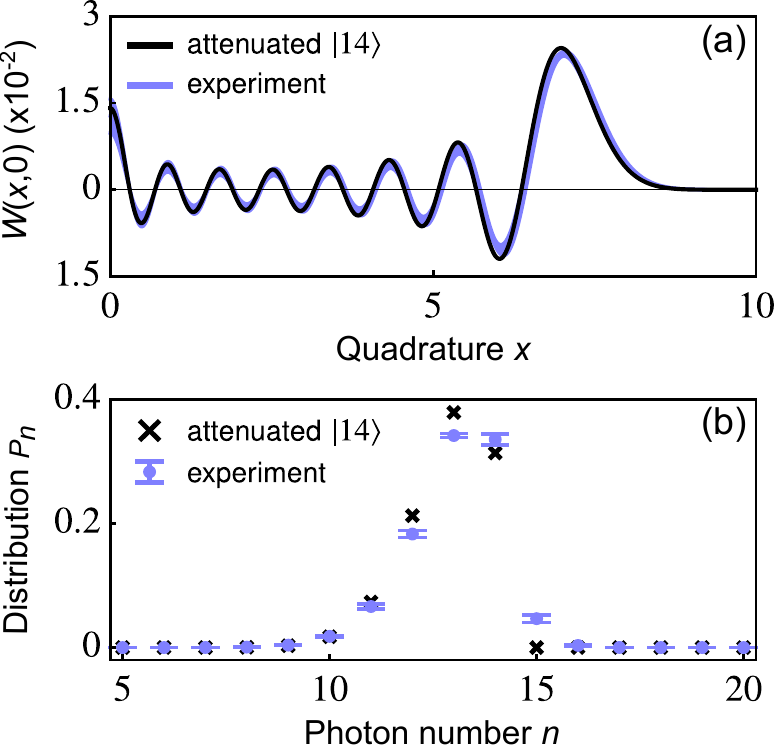}
  \caption{Quantifier output. (a) Cut through the output Wigner function of the computational quantifier with fourteen experimental photon-number statistics as inputs. The line thickness provides the $3\,\sigma$ interval for the values of the Wigner function. The black fit corresponds to the attenuated Fock state $|14\rangle$ with an attenuation $\eta=0.9205$. (b) Associated photon-number distribution (blue points) compared to the one of the attenuated Fock state $|14\rangle$ (black crosses). The output Wigner function and the photon-number distribution are averaged over thirty random choices of photon-number statistics from the data set 7, with $P_1=0.91$ and $P_{2+}=0.02$.}\label{Wigner}
\end{figure}

%intermediate states
The necessary condition for a non-trivial capability $n>1$ is to reach a single-photon component $P_1>2/3$  \cite{zapletal2017}. Above this threshold, the capability moderately grows with $P_1$. As can be seen in Fig.~\ref{cap}, the state corresponding to the green bar (set 3 in Table \ref{table}) has a multi-photon component $P_{2+}=0.02$ and the capability of the Fock state $|3\rangle$. The state associated to the red bar (set 4) has the capability of the Fock state $|4\rangle$ despite having a similar single-photon component as the previous state but a larger, still low, probability $P_{2+}=0.05$. For a given $P_1$, an increase in $P_{2+}$ may thereby lead to a larger capability. Actually, this increase in $P_{2+}$  comes in that case with a decrease in the vacuum component, indicating that the bunching is less affected by multi-photon contributions than vacuum admixture. We have shown in additional simulations that at fixed vacuum the capability decreases with the multi-photon component.

%good states
Finally, for $P_1>0.8$, the capability is expected to rapidly increase and to diverge at $P^{(\infty)}_{1}=0.885$, where an arbitrary capability can be reached \cite{zapletal2017}. The experimental results agree well with this prediction and highlight the nonlinearity of the quantifier. The verification of this trend is an important benchmark for the development of single-photon sources. The data sets indicated with blue bars have at least the capability $14$. For the set 7, note that its $g^{(2)}(0) = 0.05$ does not significantly differ from that of the states with the trivial Fock-state capability. The capability $14$ is also achieved for lower single-photon fidelities $P_1$ and higher multi-photon contributions $P_{2+}$, even for a state with four times larger $g^{(2)}(0) = 0.2$. However, these states might have a lower capability than the set 7 due to the saturation to 14 for reason of computational power.

\section{Discussion: effect of loss and truncation}
Figure~\ref{Wigner} presents the output of the computational quantifier with fourteen input states randomly chosen from the data set 7, i.e., the set with the highest heralding efficiency and lowest multi-photon component. Figure~\ref{Wigner}a first provides the cut through the Wigner function. The output Wigner function is fitted by the one of a lossy Fock state $|14\rangle$, with a fitted attenuation parameter $\eta=0.9205\pm 0.0005$. The fit shows that the oscillations of the output Wigner function in phase space coincide with the ones of the attenuated Fock state $|14\rangle$. The photon-number statistics of the output state and attenuated Fock state are compared in Fig.~\ref{Wigner}b. The good cut-off of the multi-photon contributions with more than fourteen photons in the statistics of the output state is another feature that further demonstrates the high quality of the initial single-photon states. Such result was made possible only by considering single-photon states with limited multi-photon contributions and very low vacuum admixture, as provided by the OPO-based source used in this work.

\begin{figure} [t!]
\centering
\includegraphics[width=0.85\columnwidth]{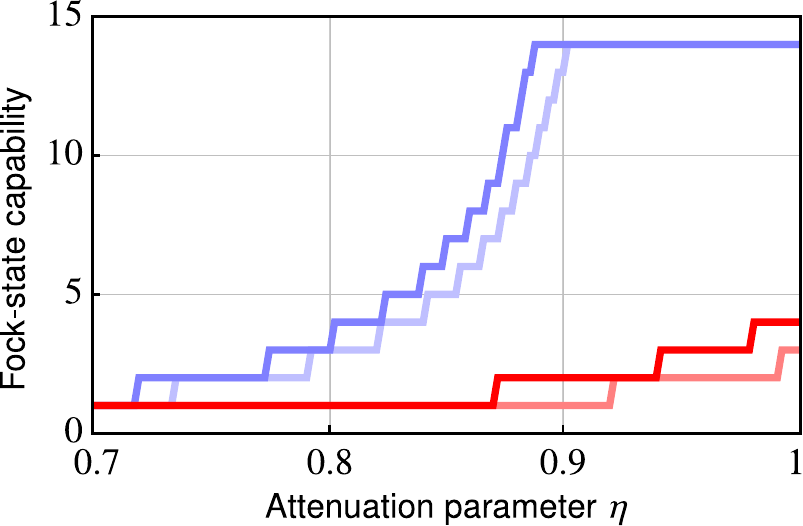}
  \caption{Fock-state bunching capability and optical loss. The blue and red lines provide the Fock-state capability for a single random choice of $n$ attenuated photon-number statistics obtained from the two data sets 7 and 4, with a capability of 14 and 4 respectively. These capabilities are compared to the capabilities of truncated states, i.e., with neglected multi-photon contributions (light blue and red lines).}\label{cut}
\end{figure}

We now come to an additional characterization of the quantifier, i.e., its evolution with optical losses. This quantifier \textit{depth}, in analogy to non-classicality depth \cite{Lee}, is tested by considering attenuation for two states randomly chosen from different data sets. Figure~\ref{cut} shows the Fock-state capability as a function of the attenuation parameter $\eta$, for the state with $P_1=0.91$ and $P_2=0.02$  (blue in Fig.~\ref{cap}) and the state with $P_1=0.74$ and $P_2=0.05$  (red in Fig.~\ref{cap}). Both states exhibit a similar $g^{(2)}(0)$ parameter (which is preserved with attenuation), but different initial capabilities 14 and 4, respectively. As it can be seen, the capability depends nonlinearly on the attenuation $\eta$. This is in contrast to the negativity of the single-photon Wigner function which decreases linearly with the attenuation. As a result, the capability allows more sensitive benchmarking of single-photon states than the negativity of the Wigner function. 

The results in Fig.~\ref{cut} are also superimposed with two plots that give the evolution of the capability with optical losses for states whose photon-number statistics are truncated, i.e., neglecting the multi-photon contribution. The discrepancy in the Fock-state capability between the experimental states and the truncated ones demonstrates that the multi-photon contributions play a significant role in such bunching experiments. The truncation of multi-photon contributions can be a limiting approximation when multi-photon interference is involved.

\section{Conclusion}
In conclusion, with the advance of quantum technologies, novel procedures and applications put challenging demands on resources and required benchmarking \cite{eisert2020}. In this broad context of utmost importance, we have employed the \textit{Fock-state bunching capability} to collectively benchmark experimental single-photon states for the first time. We have investigated the behavior of this test with photon statistics and loss. This quantifier, which is highly non-linear, has a clear operational meaning in terms of photon merging and moreover takes into account the unavoidable dispersion of individual copies of single-photon states. 

Thanks to high-purity states based on a state-of-the-art OPO, this work has experimentally verified the numerically-predicted threshold, $P_1>0.885$, to observe a large Fock-state capability. Capability of at least $14$ has been demonstrated thanks to the very low two-photon component and the large heralding efficiency. Importantly, we have shown that the capability is more sensitive to optical losses than the single-photon negativity of the Wigner function and fidelity. Based on our numerical data, we also deduced that a moderate increase in the ratio of the multi-photon contributions to the vacuum does not decrease the capability. This shows that despite the negative impact of multi-photon contributions, they prevent the bunching of single-photon states into a single Fock state less severely than optical losses.

In the present implementation, we have estimated photon-number distributions from homodyne detection. Multiplexed single-photon detectors \cite{harder2016a,bohmann2018} or photon-number resolving superconducting detectors \cite{harder2016b, klaas2018, tiedau2019} should enable a direct measurement of the Fock-state capability. Also, this benchmark does not depend on the nature of the source and can thereby be used to characterize microwave photons in superconducting circuits \cite{Pfaff2017}, plasmons at metal-dielectric surfaces \cite{Zwiller,Atwater}, phonons in trapped-ion  \cite{kienzler2017} or optomechanics experiments \cite{Hong2017}, and collective excitations in atomic ensembles \cite{laurat,ourjoumtsev,corzo}. Finally, the multi-photon interference quantifier can be modified to investigate the capability of other resource states, e.g. squeezed states or Schr\"odinger cat states \cite{Minzioni2019,Sychev2017}, to produce different target states such as NOON states  \cite{matthews2016,ulanov2016,slussarenko2017} or superpositions of squeezed states (GKP states) \cite{gottesman2001}, opening a new avenue for testing the potential of light emitters for advanced quantum state engineering.\\

\begin{acknowledgments}
This work was supported by the European Union's Horizon 2020 research and innovation framework programme under grant agreement No 731473 (QuantERA ERA-NET Cofund in Quantum Technologies, project ShoQC), No 820445 (FETFLAG Quantum Internet Alliance), and No 951737 (Twinning project NonGauss). This work was also funded by the French National Research Agency (HyLight project ANR-17-CE30-0006). R.F. acknowledges grant 20-16577S of the Czech Science Foundation and national funding from the MEYS (project 8C20002). G.G. acknowledges the support by the European Union (Marie Curie Fellowship HELIOS IF-749213), T.D. by the R\'egion Ile-de-France in the framework of DIM SIRTEQ, and P.Z. by the European Union's Horizon 2020 research and innovation framework programme under grant agreement No 732894 (FET Proactive, project HOT). We also thank K.~Huang and O.~Morin for their contributions in the early stage of the experiment.
\end{acknowledgments}

\cleardoublepage

\onecolumngrid

\section*{Appendix A: Single-photon state generation and detection}
The triply-resonant optical parametric oscillator is based on a 1-cm long type-II phase-matched KTP crystal (Raicol), pumped at 532 nm with a frequency-doubled Nd:YAG laser (InnoLight GmbH). The input face of the crystal is coated for high-reflection at 1064 nm and $R=$ 95\% for the pump, while the output curved mirror (38-mm radius of curvature) is highly reflective for the pump and has $R=$ 90\% for the infrared light. The OPO is operated well below threshold. Photon pairs emitted at 1064 nm are orthogonally polarized and separated on a polarizing beam-splitter. After frequency filtering via an interferential filter and a home-made narrow-band cavity, the idler photons are detected with a high-efficiency WSi superconducting nanowire single-photon detector \cite{LeJeannic2016}. The heralded single photons are characterized using homodyne detection \cite{maxlik}. The overall detection loss, which includes propagation losses, electronic noise, interference visibility and detector efficiency, amounts to about 15\%. 

\section*{Appendix B: Evaluation of photon statistics}
The Fock-state capability is determined for $n$ independent photon-number statistics $p_{m_j}$, where $m_j$ is the photon number and $j=1,...,n$ labels the individual statistics. We reconstruct the photon-number statistics from multiple runs of the single-photon source under the same experimental conditions. In the computer, the statistics $p_{m_j}$ conditionally merge into a single output statistics following the computational procedure described in the main text. In the first step of the procedure, the input statistics $p_{m_j}$ are mixed in a linear optics network represented by the unitary operator $\hat{U}$. To keep the quantifier unbiased, the linear optics network symmetrically mixes the inputs. In the second step, all photons are conditionally merged into a single output mode considering all the other modes in vacuum. In this way, we calculate the photon-number distribution for the output mode and the associated Wigner function. The single-photon source producing $n$ photon-number statistics $p_{m_j}$ has the capability of the Fock state $|n\rangle$ if the computed Wigner function has the same number of negative annuli as that for the ideal Fock state $|n\rangle$ \cite{schleich2001}. 

\section*{Appendix C: Capability determination: an example}
\begin{figure}[h!]
\centering
\includegraphics[width=0.66\linewidth]{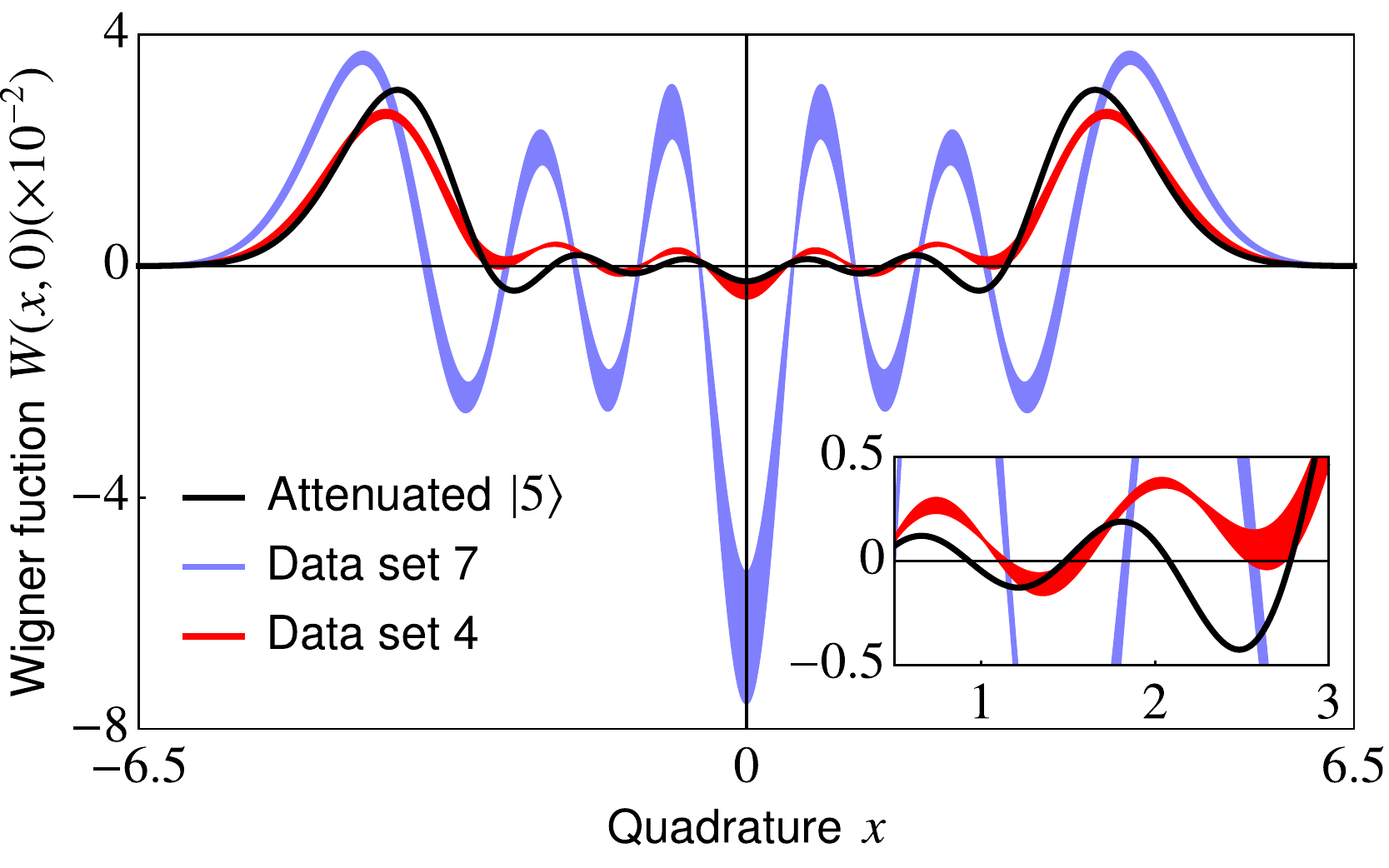}
\caption{Determining the Fock-state capability of $n$ independent non-ideal single photons. The Wigner function $W(x,0)$ of the output state of the computational quantifier for $n=5$ is averaged over 30 randomly chosen batches of $n=5$ photon number statistics that are randomly chosen from data set 4 (red line) and data set 7 (blue line). The line thickness provides the $3\sigma$ interval for the values of the Wigner function. The black line corresponds to the attenuated Fock state $|5\rangle$ with the attenuation $\eta = 0.72$.}\label{quantifier}
\end{figure}

We illustrate how the Fock-state capability is determined using the computational quantifier via a specific example. We start by taking a batch of photon number statistics randomly selected from a given data set. This batch contains a fix number $n$ of photon statistics. For concreteness we set for instance $n=5$ and consider data sets 4 and 7 (Table 1 of the main text). We apply the computational quantifier (Fig.~1 of the main text) on this batch and determine the Wigner function of the quantifier output state. The output Wigner function is averaged over 30 randomly chosen batches from a given data set and plotted in Fig.~\ref{quantifier}. The area in phase space, where the rotationally symmetric Wigner function for data set 7 is negative, forms two negative annuli and one negative circle at the origin. This can be seen in the cut through the Wigner function (blue line) plotted in Fig.~\ref{quantifier}. Following the discussion in Sec.~2 of the main text, we conclude that single photons in the data set 7 have the capability 5. On the other hand, the output Wigner function for data set 4 has only one negative annulus and one negative circle (inset of Fig.~\ref{quantifier}). We conclude that data set 4 does not have capability 5. By applying the quantifier to batches with different number $n$ of single photons we determine the maximal capability that each data set exhibits and plot this capability in Fig.~2 of the main text.

\section*{Appendix D: Computational limitations}
The data sets labeled by the blue color in Fig.~2 of the main text are expected to have a Fock-state capability higher than $14$. However, it cannot be tested due to computational limitations. The quantifier is based on the processing of multiple distinct photon-number statistics estimated from the experiment. The computational time of this processing grows exponentially with the number of input photon-number statistics. To test the Fock-state capability $n$, $m^n$ distinct terms are evaluated if one considers $m$ lowest photon components in the photon-number statistics. For practical reasons, we restrict our analysis to test the capability $14$, which is sufficient for our purpose due to the nonlinearity of the capability. This computational limitation would stay even if the projective vacuum measurement is replaced by a heterodyne measurement with post-selection. In this case, the computational limitation arises due to Gaussian integrals whose complexity scales with the number of photon-number distributions, which are on the inputs of the quantifier.

The computational demand is substantially reduced if one chooses a single photon-number distribution from the experimental data set and use it as an identical input, which is fed into all channels of the linear optics network. In this way, we calculate the output Wigner function for several random choices of the photon-number distribution. Then the Wigner function is averaged over these random choices. Hence the differences between the individual copies of single-photon states are not taken into account. Using this simplified method, we determine the Fock-state capabilities of the experimental data sets, which agree with the capabilities depicted in Fig.~2 obtained by the full, unsimplified multi-photon interference quantifier. However, note that the full quantifier should always be used to confirm the results of the simplified quantifier, which is not able to correctly estimate the propagation of the input state's discrepancies through the quantifier. In order to estimate the capability of a very high Fock state, the quantifier can be even further simplified by neglecting the discrepancies between photon-number distributions in the experimental data set. Working only with the average photon-number distribution, we estimate that the experimental data set 7 with $P_1=0.91$ and $P_2=0.02$  has the capability of at least the Fock state $|50\rangle$. This agrees with the theoretical prediction \cite{zapletal2017} that $P_1>P_1^{(\infty)}=0.885$ is sufficient to reach the capability of an arbitrary Fock state, if multi-photon contributions and discrepancies between photon-number distributions are neglected.

\end{document}